# МОДУЛЬНАЯ ТЕХНОЛОГИЯ РАЗРАБОТКИ РАСШИРЕНИЙ САПР. АКСОНОМЕТРИЧЕСКИЕ СХЕМЫ ТРУБОПРОВОДОВ. ТИПОВЫЕ И СПЕЦИАЛЬНЫЕ ОПЕРАЦИИ

Сафин И.Т., Мигунов В.В., Кафиятуллов Р.Р.
ЦЭСИ РТ, Казань
vmigunov@csp.kazan.ru

Рассматривается применение модульной технологии разработки проблемно-ориентированных расширений САПР к задаче автоматизации подготовки аксонометрических схем трубопроводных систем на примере программной системы TechnoCAD GlassX. Обсуждаются особенности реализации типовых операций, состав и реализация специальных операций проектирования схем специальных технологических трубопроводов, систем водопровода и канализации, отопления, теплоснабжения, вентиляции, кондиционирования воздуха.

Настоящая работа посвящена применению модульной технологии разработки проблемно-ориентированных расширений САПР реконструкции предприятия, общие положения которой изложены в [1]. Объект приложения технологии - автоматизация подготовки аксонометрических схем трубопроводных систем (АСТС), выполненная в САПР TechnoCAD GlassX. Параметрическое представление АСТС (ПП), разработанное в рамках модульной технологии, изложено в [2]. Здесь рассматриваются особенности реализации типовых операций над этим ПП, а также состав и реализация специальных операций.

Для реализации всех операций имеется основное меню "Аксонометрическая схема" (рис.1), из которого открываются другие меню, более подробные. Сильно разветвлены опции основного меню: "Добавление" (рис.2), "Редактирование" (рис.4), "Установки" (рис.5), "Схему в чертеж" (рис.3). Особенно много возможностей предоставляет "Редактирование", где после указания видимого объекта предлагается

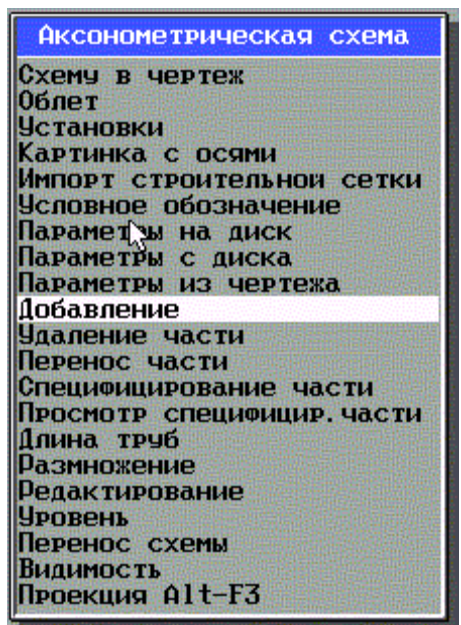 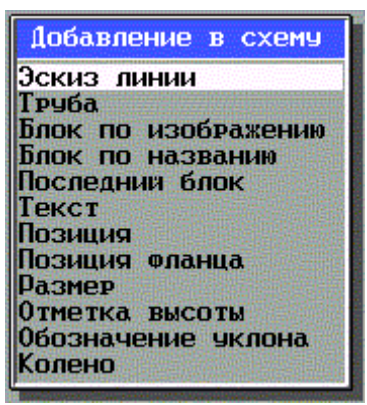

Рис.1. Основное меню         Рис.2. Операции добавления объектов выбрать какую-либо операцию по его изменению. В этих меню типовые операции технологии никак не отделяются от специальных.

*Типовые операции*

Из особенностей реализации типовых операций следует отметить группу операций по помещению схемы в чертеж (рис.3), нестандартные операции по добавлению сразу нескольких труб "Эскиз линии" и "Колено", простановку группового позиционного обозначения фланцевого соединения, нанесение условных обозначений (блоков) на трубы, простановку цепных размеров, помещение операций удаления и изменения свойств объектов в режим их редактирования, в котором показываются все опции, но доступны только возможные для выбранного объекта (рис.4), операцию удаления труб.

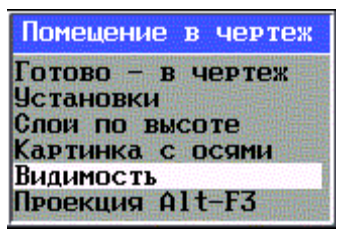
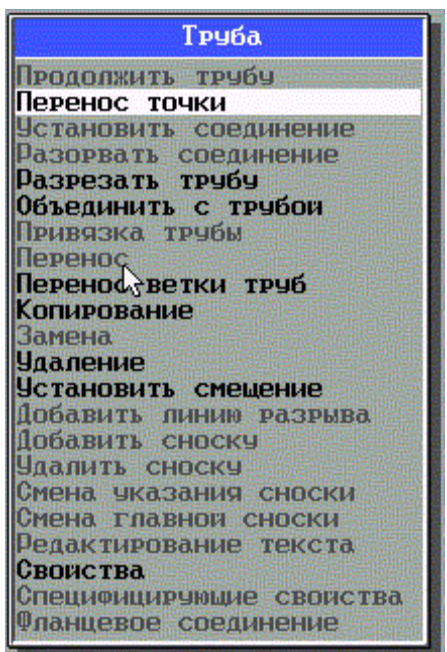

Рис.3. Операции группы "Схему в чертеж"     Рис.4. Редактирование труб

При помещении схемы в чертеж предоставляется возможность поменять установки, действующие на АСТС (рис.5), выбрать слой по высоте (этаж), установить видимость объектов схемы (она могла быть неполной для удобства во время работы), нужную аксонометрическую проекцию. Большинство этих операций доступны из основного меню, но при помещении в чертеж наличие этих опций служит напоминанием о том, что соответствующие параметры должны быть заданы корректно перед помещением результатов в чертеж. Здесь же предлагается сгенерировать картинку, задающую направления осей координат для помещения в чертеж вместе со схемой. Во время работы по подготовке схемы эта картинка не используется и может появиться только здесь.

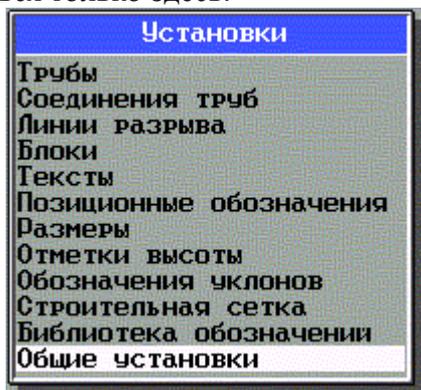

Рис.5. Меню установок АСТС

Операция "Эскиз линии" реализует быстрое нанесение линии трубопровода с трубами, идущими только вдоль осей координат. Сначала строится обычная ломаная с ортогонализацией по текущей аксонометрической проекции. При этом возможны привязки к имеющимся в схеме точкам. Затем автоматически все отрезки ломаной превращаются в соединенные трубы. Концы линии с остальными трубами схемы не соединяются.

Операция "Колено" – врезка в трубу колена для температурной компенсации или других целей. Запрашивается выбор трубы, затем точек начала и конца колена на ней. Положение точек вдоль трубы можно уточнять путем ввода чисел. Затем курсором указывается сдвиг колена, который также можно уточнить. При указании сдвига одновременно выбирается и его направление – координатная ось. Это достигается за счет включения режим ортогонализации. Курсор от конца колена может идти только по осям координат.

"Позиция фланца" – простановка группового позиционного обозначения с несколькими (в соответствии с установкой) позиционными номерами на блок. Позиции задаются вручную или нумеруются автоматически, начиная со следующего за последним занятым номером, в зависимости от соответствующей установки. Для группового позиционного обозначения с 4-мя или 5-ю позициями допускается автоматизированный последовательный выбор ряда параметров, приводящий к одновременному назначению всех требуемых специфицирующих свойств комплекта деталей фланцевого соединения. При этом 1-я позиция соответствует видимому на схеме блоку, 2-я – невидимым шпилькам (болтам), 3-я – невидимым гайкам, 4-я – невидимым шайбам, 5-я (если есть) – невидимым прокладкам (линзам). Это позволяет поднять степень автоматизации работ при специфицировании фланцевых соединений, которое необходимо для специальных технологических трубопроводов (высокого давления) и в ряде других случаев.

При помещении условных графических обозначений трубопроводной арматуры и элементов трубопроводов в АСТС (рис.2, "Блок по изображению", "Блок по названию", "Последний блок") они выбираются в текущей подключенной графической библиотеке и затем "привязываются" к трубам. Если в окрестности курсора находится конец трубы – блок привязывается к этому концу. Пространственная ориентация блока изменяется путем нажатия клавиши пробел по всем возможным вариантам с учетом требования, что через его плоскость требуется провести выносную линию вдоль одной из осей координат. Если блок допускает угловую или тройниковую привязку и установлен на конец трубы, из которого выходят другие трубы – запрашиваются еще вторая (для тройниковой – и третья) соединяемая труба. После их указания делается попытка их привязки. В случае, если привязка удается (трубы идут в нужных направлениях), все трубы, к которым привязан блок, соединяются между собой попарно. Их части, попадающие под блок, пропадают. Если труба не закрыта блоком, а видна сквозь него, значит, привязки блока к этой трубе нет. Если блок привязан только к одной трубе, запрашивается уточнение его позиции вдоль трубы.

При нанесении цепных размеров на схеме появляются изображения точек крестиками, и запрашивается выбор точек начала выносных – концов труб и точек привязки блоков. Затем запрашивается положение размерной линии, указываемое курсором с подсветкой варианта размера. Переключение вариантов среди допустимых для указанных точек – клавишей пробел. При этом обеспечиваются только такие направления размерной линии и выносных, которые удовлетворяют требованиям действующих стандартов, подробно изложенным применительно к размерам в АСТС в [2].

При удалении трубы связанные с ней соединения, блоки, сноски и отметки высоты также будут удалены. Также будут удалены тексты и позиционные обозначения, все сноски которых указывали на эту трубу и ее блоки, и размеры, у которых остается меньше двух точек начала выносных. Если труба соединяла другие трубы с местными смещениями, их смещения могут измениться – они сдвинутся на схеме.

*Специальные операции. Основное меню*

Переходя к специальным операциям, начнем с операции "Облет", расположенной в основном меню рядом с помещением схемы в чертеж. Реализуется последовательный просмотр схемы с различных позиций для получения полного наглядного представления. Задается число шагов, начальное положение и угловой шаг – и схема начинает поворачиваться на каждое нажатие клавиши Enter (рис..6)

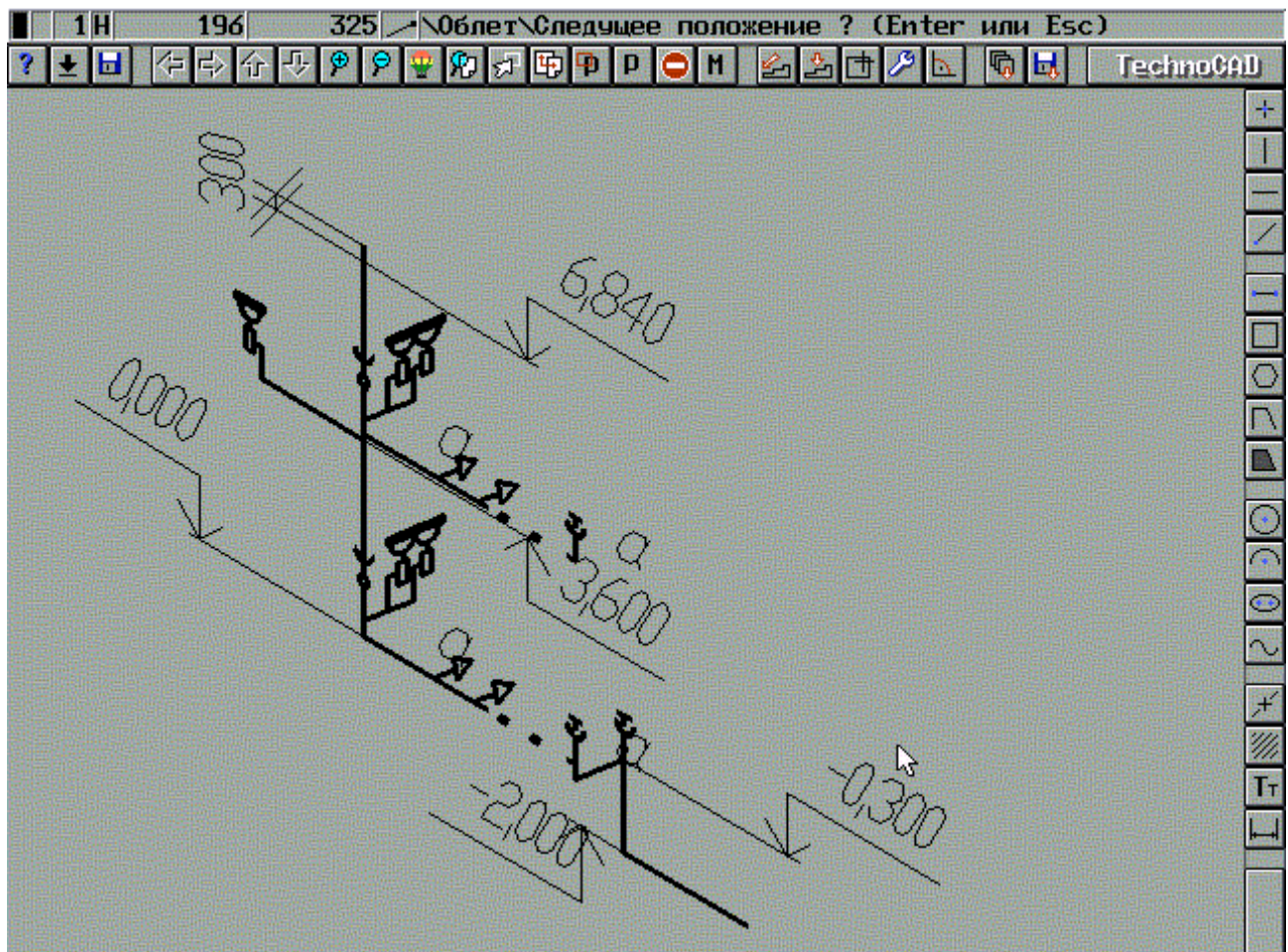

Рис.6. "Облет" АСТС с произвольными угловыми шагами в пространстве

Специальная операция "Импорт строительной сетки" – ввод в схему данных по сетке строительных осей из плана подосновы в чертеже либо из комплекта параметров плана подосновы на диске. Здесь достигается высокий уровень автоматизации ввода данных и интеграция с другим расширением TechnoCAD GlassX.

Для помещения в АСТС условных графических обозначений используется выбор этих обозначений в специальной графической библиотеке. Операция "Условное обозначение" – создание нового условного графического обозначения арматуры или элементов трубопроводных систем. Это обозначение затем можно будет поместить в графическую библиотеку и наносить на аксонометрические схемы. Обозначение для помещения на схему должно состоять из одного геометрического элемента типа модуль и иметь свойства, позволяющие его автоматически привязать к трубопроводу и правильно определить вырезаемую на трубах часть. Привязка должна быть осевой, угловой или тройниковой, но не точечной. В случае осевой привязки дополнительно можно задать вырезаемый на трубе отрезок. Если привязка угловая или тройниковая – исключаемая часть труб определяется самими точками, заданными в привязке. Допустимые геометрические элементы обозначения: отрезки, прямоугольники, ломаные, закрашенные ломаные, окружности, дуги окружностей, многоугольники - их можно поворачивать и растягивать их в трехмерном пространстве. По данной опции программа будет контролировать все эти требования. Кроме перечисленных свойств обозначения, желательно указывать наличие симметрии относительно оси привязки и нормали к ней геометрии обозначения и его привязок – это сокращает число вариантов различной пространственной ориентации блока, которые будут перебираться пользователем при установке блока на трубы.

"Удаление части" – предлагается выбрать в схеме множество объектов и убрать их из схемы. Удаляемое множество расширяется автоматически до такого состояния, когда остающиеся объекты не будут иметь ссылок на удаляемые объекты. Например, если отмечена к удалению труба, то все блоки, установленные на нее (и только на нее) будут также удалены. Ввиду вероятности потери информации предоставляется возможность отказаться от удаления после просмотра результатов.

"Перенос части" – предлагается выбрать множество труб в схеме, затем задать сдвиг при переносе. Переносится все, что связано с этими трубами – блоки, отметки высоты, размеры, сноски, тексты.

"Специфицирование части" – загрузка редактора таблиц для правки специфицирующих свойств элементов с выбранными позиционными обозначениями. При этом обеспечивается доступ к электронным каталогам (рис.7).

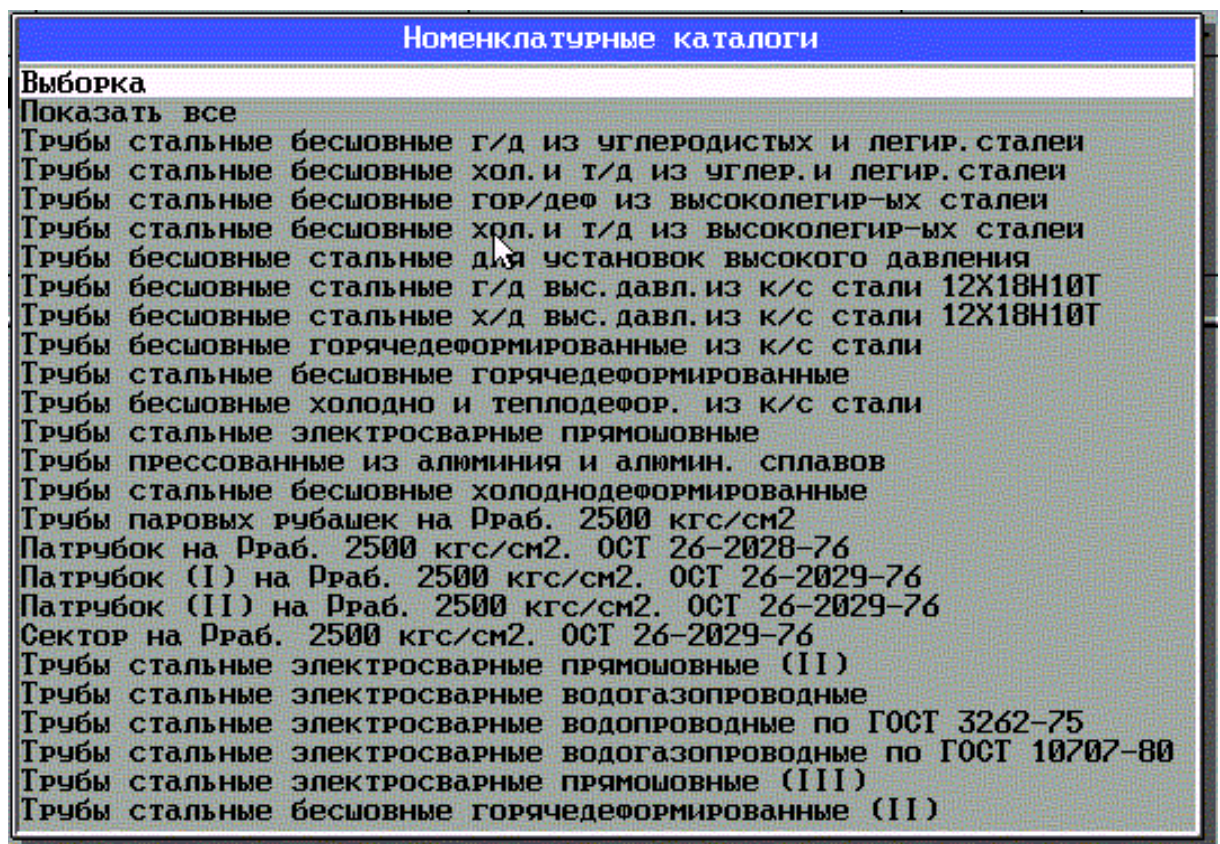

Рис.7. Выбор из списка имеющихся каталогов труб

"Просмотр специфицир.части" – просмотр выделенных цветом труб и блоков, которые подготовлены для специфицирования. Циклическое переключение между просмотром специфицируемой части (элементов, имеющих позиционные обозначения) и "неназначенной" части – элементов с позиционными обозначениями, среди позиционных номеров которых имеются такие, что в соответствующих специфицирующих свойствах не назначено ни одного свойства кроме позиции и количества. Наличие в схеме "неназначенной" части элементов говорит о необходимости задания им специфицирующих свойств до перехода к генерации спецификации. Выделение специфицируемой части позволяет определить, правильно ли задано на схеме все, что нужно специфицировать.

"Длина труб" - режим выбора множества труб, чья суммарная длина в ходе выбора показывается в информационной строке вверху экрана. Операция полезна, когда специфицируется часть схемы.

"Размножение" – создание одной или нескольких копий выбираемого в схеме множества. Если выбран один блок – его предлагается тиражировать только вдоль трубы, на которую он установлен. Если что-либо иное – сдвиг на каждом шаге и число копий могут быть произвольными. Проводится проверка множества на допустимость копирования, затем осуществляется размножение в чертеже цветом построений и предлагается просмотреть результат. В ходе просмотра можно менять текущую проекцию схемы.

*Специальные операции редактирования (рис.4)*

"Продолжить трубу" – опция доступна, если был указан конец трубы, не соединенный с другими трубами. Указывается точка вдоль оси трубы, куда должен сместиться выбранный конец трубы, затем его новая позиция уточняется в форме ввода. Связанные с трубой блоки, сноски и отметки высоты будут переноситься в следующих случаях:

- продолжаемый конец трубы оказался с другой стороны от неподвижного конца трубы – блоки сместятся симметрично относительно неподвижного конца трубы;
- труба укоротилась, и блоки перестали на ней умещаться – блоки сместятся к ближайшему концу трубы.

"Перенос точки" – если к указанному концу трубы сходятся другие трубы, с которыми он не соединен, то программа задаст вопрос, переносить точку для всех труб или только для выбранной трубы. Если выбранная труба соединена с другой трубой, всегда переносится точка для всех труб. Указывается новая точка со смещением от старой только вдоль осей текущей проекции. Ее координаты уточняются в форме ввода. Объекты, связанные с трубой, поведут себя так же, как при операции "Продолжить трубу".

"Установить соединение" – опция доступна, если был указан конец трубы, который можно соединить с другой трубой. Соединить две трубы можно в том случае, если их концы находятся в одной точке и соединение на них еще не установлено. Для соединения выбирается одна из труб, допускающих соединение с указанной. Обратная операция - "Разорвать соединение".

"Разрезать трубу" – указывается и затем уточняется точка разреза на трубе. Точка разреза может оказаться на конце другой трубы, тогда части разрезаемой трубы включаются в ее ветки, и на них начинают действовать местные смещения этой ветки. Части трубы могут сместиться на схеме.

"Объединить с трубой" – опция доступна, если имеется хоть одна труба, продолжающая выбранную в одну из сторон, и сращивание этих труб возможно, т.е. точка сращивания не является концом выносной для какого-либо размера и точкой привязки для блока, соединенного с другой трубой. Если объединение возможно с двумя трубами, то выбирается точка сращивания указанием на схеме. При объединении двух труб другие трубы, отходящие от их точки сращивания, выходят из состава некоторых веток, в которые они входили. На них перестают действовать соответствующие местные смещения, и они могут переместиться в схеме.

"Привязка трубы" – опция доступна, если использованы не все привязки блока. Соединит блок с еще одной трубой. Труба выбирается указанием в схеме. Блок может быть соединен с трубой, если пространственный угол между его лучом привязки и осью трубы не более 45°. Если привязываемая труба не лежит в плоскости блока, он поворачивается в плоскость, образуемую трубами.

"Перенос ветки труб" – опция доступна, если указана труба, имеющая соединения. Автоматически генерируется и подсвечивается множество труб (вместе со связанными с ними объектами), соединенных с указанной посредством соединений и других труб, после чего дается запрос о продолжении операции над собранным множеством. Затем указывается новое положение ветки труб со смещением только вдоль осей текущей проекции. Смещение уточняется в форме ввода. Ориентация ветки не изменяется. Объекты, связанные с трубами, не меняют своего положения на трубах.

"Замена" – заменяет указанный блок другим блоком из библиотеки с выбором по изображению, по названию либо последним выбранным. При замене на блок с меньшим количеством привязок (например, блока с тройниковой привязкой на блок с угловой или осевой привязкой) возможна потеря части привязок.

"Установить смещение" – выбирается тип смещения: общее или местное. Затем последовательно запрашиваются граница смещаемой области (точка на трубе, задающая перпендикулярную к трубе плоскость) и что будет смещаться (вторая точка на трубе, положение которой относительно плоскости задает смещаемое полупространство). Для общего смещения на этом операция завершается, а для местного из предложенных для выбора труб предлагается выбрать разрываемые трубы, указывая на них позицию добавляемых линий разрыва.

"Смена указания сноски" – запросит новую точку указания для сноски на трубе или на блоке.

"Смена главной сноски" – опция доступна, если текст имеет более одной сноски. Запросит указание неглавной сноски и сделает ее главной.

"Специфицирующие свойства" – загрузка редактора таблиц для правки специфицирующих свойств трубы или блока, если у них есть позиционное обозначение. Свойство "Количество" при этом игнорируется. Имеется возможность временного выхода в АСТС для просмотра всех элементов, которых касается текущий позиционный номер.

"Фланцевое соединение" – для позиционного обозначения с 4-мя или 5-ю позициями выполняется последовательность выбора ряда параметров, приводящая к одновременному назначению всех требуемых специфицирующих свойств комплекта деталей фланцевого соединения.

*Специальные операции. Основное меню. Продолжение*

"Уровень" – предлагается выбрать отметку высоты и указать ее новый уровень. Все точки в схеме соответственно поднимутся или опустятся (не на экране), и все отметки высоты получат новые значения.

"Перенос схемы" – перенос всей схемы в другое место на поле чертежа.

"Видимость" – установка видимости объектов схемы. Позволяет во время работы оставлять в схеме только нужные элементы, не загромождая ее.

"Проекция Alt-F3" – изменение текущей проекции схемы. Эта операция доступна также в большинстве режимов просмотра АСТС и выбора ее объектов по

"горячей клавише".

Специальной также является операция изменения текущей подключенной графической библиотеки обозначений в группе операций "Установки".

Следует отметить одну особенность пользовательского интерфейса в операциях редактирования. После выбора в основном меню этого раздела сначала запрашивается выбор объекта, а затем предлагается выбор конкретной операции. В остальных случаях (не в разделе "Редактирование") первичен выбор операции, и в ходе выбора объекта сразу подключается режим доступности только тех объектов, которые допускают выбранную операцию. Дело в том, что в случае назначения отдельных опций для каждой операции редактирования каждого типа объектов меню редактирования не поместилось бы на экран, и пришлось бы вводить многоуровневые меню выбора нужной операции, что замедляет работу проектировщика.

Изложенные в настоящей работе состав и особенности реализации операций подготовки АСТС вместе с их структурированным параметрическим представлением [2] являются системной моделью аксонометрических схем трубопроводных систем, используемых при проектировании реконструкции предприятий согласно требованиям стандартов системы проектной документации для строительства. Модель первоначально разработана в 1999 году по модульной технологии [1] и в дальнейшем проявила высокую степень устойчивости к добавлению новых возможностей. За четыре года в рамках тех же модельных представлений в подготовку АСТС были добавлены возможности специфицирования трубопроводов обычного [3] и высокого давления [4], чертежей марок ОВ, ВК, НВК, ТС. Дополнения вносились практически без изменения уже имевшихся списков объектов и операций. Это говорит как о высокой степени адекватности параметрического представления и списка операций объективной сущности АСТС, так и об эффективности модульной технологии разработки проблемно-ориентированных расширений САПР реконструкции предприятия.

*Использованные источники*